\documentclass[conference]{IEEEtran}
\IEEEoverridecommandlockouts
\usepackage{cite}
\usepackage{amsmath,amssymb,amsfonts}
\usepackage{algorithm}
\usepackage[noend]{algpseudocode}
\usepackage{graphicx,epstopdf}
\usepackage{textcomp}
\usepackage{xcolor}
\usepackage{blindtext}
\usepackage{multirow}

\def\BibTeX{{\rm B\kern-.05em{\sc i\kern-.025em b}\kern-.08em
    T\kern-.1667em\lower.7ex\hbox{E}\kern-.125emX}}
\begin{document}

\title{Relative Positioning of Autonomous Systems using Signals of Opportunity}
\author{Nicolas Souli, Panayiotis Kolios, and Georgios Ellinas % <-this % stops a space
\thanks{Nicolas Souli, Panayiotis Kolios, and Georgios Ellinas are with the Department of Electrical and Computer Engineering and the KIOS Research and Innovation Center of Excellence, University of Cyprus, {\tt\small \{nsouli02, pkolios, gellinas\}@ucy.ac.cy}}}

\maketitle

\begin{abstract}
For reliable operation, next generation autonomous agents will need enhanced situational perception as well as precise navigation capabilities. The global navigation satellite system (GNSS) signals that are utilized by  practically all modern positioning systems cannot satisfy this requirement for heighten autonomy levels and positioning is becoming a decisive factor for their proliferation. 

This work investigates how relative positioning can be achieved using signals that are already accessible in the environment, and derives an online procedure for the exploitation of these signals for localization in GNSS-challenged areas. The proposed relative positioning system (RPS) explores the signal properties over a large spectrum of frequency bands, and derives a vehicle tracking algorithm to accurately estimate the vehicle's trajectory in space and time using an arbitrary set of unknown reference positions. Experimental results demonstrate the applicability of RPS and investigate its performance over the different parameter values.
\end{abstract}
\begin{IEEEkeywords}
Signals of opportunity, relative positioning, vehicle tracking
\end{IEEEkeywords}

\section{Introduction}
As vehicular technology advances towards higher levels of automation, a stable and precise navigation system is becoming a mandate for safe and effective operation \cite{Maaref2018a}. Currently, navigation systems primarily rely on GNSS for localization purposes \cite{Simkovits2017}. However, GNSS signals frequently become unreliable (e.g., in the presence of interference, jamming, or spoofing \cite{Shamaei2018a, Kassas2017, Michel2018} and in deep urban canyons \cite{Kapoor2017, Cooper2015, Morales}). In turn, navigation systems utilize an inertial navigation system (INS) \cite{Morales2018b}, and light and range sensors \cite{Maaref2018} to estimate their location in the absence of GNSS signals. The shortfall of these solutions, including data degradation, loss of signal due to multipath and antenna obstruction, have led to the exploration of alternative methods for localization \cite{Kapoor2017, Shamaei2018}. 

Lately, the utilization of signals of opportunity (SOPs)  has been proposed as an attractive alternative for navigation, when GNSS signals become unreliable \cite{Morales,Maaref2018a,Raquet2007}. SOPs (AM/FM radio, cellular, TV signals, etc.) are appealing for navigation purposes because these signals are readily available and are received at high power~\cite{Raquet2013,Morales2018b}. Hence, a number of previous studies have looked into utilizing SOPs, taking into consideration a priori knowledge regarding the receiver's reference position and the transmitter's location, or fused information using GNSS, in an effort to address the navigation shortfalls in areas where GNSS signals become unreliable. 

Complementary to those approaches, this work proposes the utilization of SOPs for relative positioning and demonstrates how this can be achieved in an online fashion. Overall, the contributions of this work are summarized as follows:
\begin{itemize}
   \item An innovative relative positioning system is proposed without the need of any a priori information about the transmitters' location or any GNSS information.
   \item An extended Kalman filter (EKF) is integrated in the proposed approach in order to achieve better localization performance.
   \item The proposed system has been experimentally implemented and thoroughly evaluated and validated in an outdoor environment.
\end{itemize}

The rest of the paper is structured as follows. Related work is included in Section~\ref{related} and the description of the methodology followed is included in Section~\ref{methodology}. Section~\ref{framework} elaborates on the proposed framework of the current work, while the experimental results are presented in Section~\ref{results}. Finally, the main conclusions regarding this work are presented in Section~\ref{conclusions}.  

\section{Related Work}\label{related}
Various works in the literature have focused on developing navigation alternatives and have tried to perform localization using SOPs. In an effort to achieve relative positioning, the utilization of RSS measurements in an indoor environment has been proposed as a cost-effective solution \cite{Silva2016}. Specifically, a weighted centroid method was derived, exhibiting better performance compared to classical fingerprinting. Nevertheless, RSS techniques were shown to be prone to interference and unreliable \cite{Silva2016}, \cite{Ma2015}. Thus, filtering techniques such as Kalman filtering \cite{Brown2011,Sazdovski2005,Liu2016,Yang2015a} and particle swarm optimization techniques \cite{Liu2016} have been utilized to improve on the RSS approach \cite{Ma2015,Li2018,Ge2015}. Additionally, techniques where GPS and SOPs are combined have been presented in order to improve navigation \cite{Simkovits2017}.

Further, a recent study in \cite{Yang2015a} has presented a tracking and positioning technique that uses mixed signals. This work utilized the time of departure (TOD) from a source together with an initial known position to achieve relative positioning. Additionally, in \cite{Kassas2014a}, GPS signals and downmixed CDMA signals are used in an adaptive estimation of a moving vehicle using SOPs. Another effort for relative positioning, proposed in \cite{Liu2016}, considers both the signal strength and range information fused in a probabilistic framework leading to more accurate positioning. 

Finally, in \cite{Wickert2018}, GPS pseudorange measurements have been utilized for localization, while in \cite{Martin2009a} a navigation system was introduced that does not rely on GPS and uses time difference of arrival (TDOA) for localization based on SOPs. Cellular signals fused with GPS and other sensor information have also been utilized in an effort to provide a better navigation method for UAVs \cite{Kassas2017}. This approach followed a TOA-based approach to achieve multisignal navigation, showing that the positioning improved compared to GPS.

Contrary to the aforementioned studies, the proposed system presents an innovative approach for relative positioning that does not take into consideration the location of the transmitters, it does not use any GNSS signal information for the relative trajectory extraction, and is utilizing only the SOPs information in an outdoor environment for localization without the use of any other sensor information. In addition, using relative ranging an EKF approach is also utilized in an effort to improve the performance of the proposed framework and achieve better relative positioning.

\section{Methodology}\label{methodology}

\subsection{System Model}
\begin{figure}[h]
\begin{center} 
 \setlength\abovecaptionskip{-0.3\baselineskip}
 \setlength\belowcaptionskip{-0.4 pt}                                                        
\includegraphics[width=9cm]{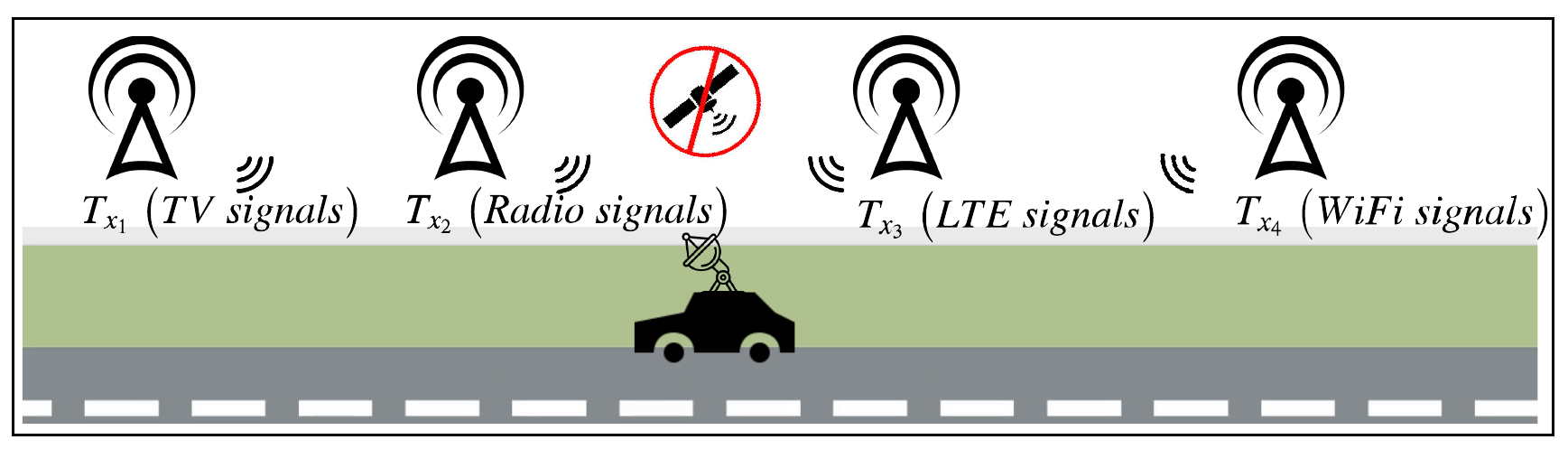}
\caption{System model.}
    \label{fig:model}
\end{center}
\end{figure}

Figure~\ref{fig:model} illustrates the components of the proposed system. A software defined radio (SDR) receiver is installed in a moving vehicle gathering SOPs, while following a route and localizes itself without any GNSS information.  

\setlength{\textfloatsep}{-2pt}

A vehicle is assumed to navigate in an environment where several SOP transmitters are present. The location of the transmitters is assumed to be unknown and a reference coordinate is used as the initial position of the receiver, calculated using the first set of data obtained from those transmitters. The various components of the system model are described below. 

\subsection{SOP Dynamic Model}
A SOP originates from a relative stationary transmitter $(T_{x})$, with a state vector $x_{s_{k}}$ consisting of its planar state $ x_{s_{k}} =[x \: \: \:  y] ^{^{T}}$ and having a discretized model $x_{s_{k+1}}= \mathbf{F}x_{s_{k}}+\mathbf{w}$, where $\mathbf{F}$ denotes the system dynamics and $\mathbf{w}$ denotes the process noise.

\subsection{Vehicle Dynamic Model}
It is assumed that the vehicle's position and velocity evolve based on a known linear model. The vehicle's state vector at time $k$ is defined as $x_{k}=[r \: \: \; \dot{r}]$, where $r=[x_{v}\: \: \; y_{v}]^{T}$ is the planar position of the moving vehicle and $\dot{r}$ is its velocity. The discrete vehicle dynamics at sampling time $T_{S}$ are denoted as $x_{k+1}= \mathbf{F}x_{k}+\mathbf{w}$.

\subsection{RSS Filtering}
The RSS measurements of the SOPs  present notable variances due to multipath and fading phenomena and could lead to inaccurate distance calculations and signal propagation model deterioration. This fact has led to the need for RSS measurements filtering using a smoothing filter to minimize the significant variations and achieve higher accuracy. Details regarding the developed filtering approach are described in the next section.
 
\subsection{Signal Propagation Model}
Standard path-loss modeling is employed to estimate the distance between transmitter $(T_{x})$ and the receiver, which is subsequently used to express the localization problem in a two-dimensional space similar to the works in \cite{Ge2015,Li2018,Tomic2018,rappaport}. 

Let $\boldsymbol{x}\epsilon \mathbb{R}$ be the unknown location of the SDR receiver and $\boldsymbol{q}_{i}\epsilon \mathbb{R}$ the unknown location of the $i$-th base station ($T_{x_{i}}$), with $i=1,...,n$. It is assumed that the distance estimates are derived from the RSS data exclusively, as the RSS approach does not impose any additional hardware demands. The log-distance path-loss model is defined as:
\begin{equation}
P_{L}=P_{T_{dBm}}-P_{R_{dBm}}
\label{pathloss}
\end{equation}

\begin{equation}
P_{L}=P_{L0} - 10n_{PL} \mathrm{log}_{10}(\frac{d}{d_{0}})+X_{G}
\label{pathloss2}
\end{equation}

\noindent where, $P_{L}$ is the total path loss, $P_{T_{dBm}}$ and $P_{R_{dBm}}$ are the transmit and receive powers in $dBm$, $P_{L0}$ is the RSS at a reference distance $d_{0}$ ($1$ meter in this case), $n_{PL}$ is the path loss exponent, and $X_{G}$ is the log-normal shadowing term modeled as a normal Gaussian variable with zero mean. The term $P_{L0}$ can be calculated using the free space path-loss model as follows: 
\begin{equation}
P_{L0}=20\mathrm{log}_{10}(d)+20\mathrm{log}_{10}(f_{c})-27.55
\label{PL0}
\end{equation}

In order to calculate the estimated distances between the base stations and the SDR receiver the log-distance path loss model can be re-written as:
\begin{equation}
d=d_{0}10^{\frac{P_{L0}-P_{L}}{10n_{PL}}}
\label{eqn:PL}
\end{equation}

The log-distance path loss model suffers from multipath and fading phenomena affecting the RSS values. In addition, environmental conditions govern the path loss exponent $n_{PL}$ which takes different values for different environments (e.g., in urban areas $n_{PL}$ values range from  $2.7$ to $3.5$) and influences the distance calculation. 

\subsection{Multilateration Method}
The distance estimation of the receiver-transmitter link as elaborated above is necessary in order to calculate the best position estimation utilizing a multilateration method. Multilateration is a conventional technique to compute the unknown location of a receiver node as discussed in \cite{Wang2013,Tomic2018,Silva2016}. At least four transmitter nodes are required for this method and the distances between these nodes and the receiver node must be calculated. For this method, it is assumed that each transmitter node transmits the information in a circle around itself with the radius of the circle being the distance to the receiver node. The location of the unknown node can be found at the intersection of these circles. The following formula can express the distance in a plane from each transmitter to the mobile node:
\begin{equation}
d_{n}=\sqrt{(x_{R_{x}}-x_{T_{x}})^{2}+(y_{R_{x}}-y_{T_{x}})^{2}}
\label{dn}
\end{equation}

When the estimated distances from the SOP transmitters to the mobile receiver are calculated using the log-normal distance model, an overdetermined system is created that does not have a unique solution. By subtracting one equation from the others, one by one, the system will be linearized in the form of $Ax_{k}=b$. Having a linearized form of the system with:
\begin{equation}
A=2\begin{bmatrix}
(x_1-x_2) & (y_1-y_2)\\ 
...&... \\ 
 (x_1-x_n)& (y_1-y_n) \\
 
\end{bmatrix}\\
\\
\label{A}
\end{equation}

\begin{equation}\label{b}
b=\begin{bmatrix}
x_1^2-x_2^2+y_1^2-y_2^2+d_2^2-d_1^2\\ 
...\\ 
x_1^2-x_n^2+y_1^2-y_n^2+d_n^2-d_1^2\\ 
\end{bmatrix}
\end{equation}

\begin{equation}\label{xk}
x_{k}=\begin{bmatrix}
x\\ 

y\end{bmatrix}
\end{equation}
the overdetermined system can be solved using the least square (LSQ) method \cite{Brown2011,Chenshu2003}, that provides the best approximated solution of the system:
\begin{equation}
e=Ax_{k}-b
\label{e}
\end{equation}
\begin{equation}
\widehat{\mathbf{x}}^{*}= \mathrm{argmin}_{{x_{k}}}e
\label{argminxk}
\end{equation}
\begin{equation}
\widehat{\mathbf{x}}^{*} = (A^{T}A)^{-1}(A^{T}b)
\label{eqn:LSQ}
\end{equation}

\subsection{EKF Localization}
\begin{figure}[h]
\begin{center}
 \setlength\abovecaptionskip{-0.6\baselineskip}
 \setlength\belowcaptionskip{-0.4 pt}
\includegraphics[width=9cm]{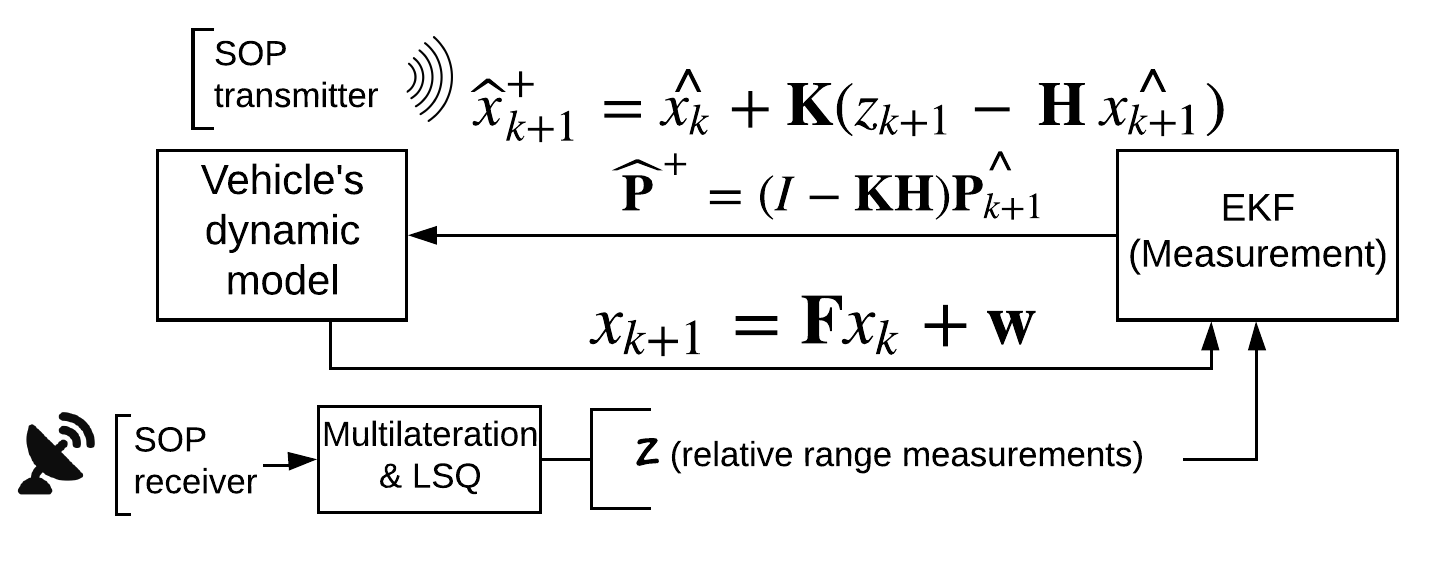}
\caption{EKF architecture.}
    \label{fig:EKF}
\end{center}
\end{figure}

The last component of the system model is the extended Kalman Filter (EKF) in an attempt to produce optimized relative coordinates $(\widehat{{X}^{*}}$). Using the data for the state of the vehicle at time $k$ in order to calculate the state at time $k+1$, the EKF can predict the vehicle's position and create an estimated trajectory~\cite{Brown2011,Wickert2018,Sazdovski2005,Li2018}. The first step, prediction, includes the $\widehat{x}_{k+1}$ and error covariance  $\widehat{\mathbf{P}}_{k+1}$ estimate. This is followed by the measurement update step, where the Kalman gain $\mathbf{K}$ and the next error covariance matrix $\mathbf{\widehat{P}^{+}}$ along with the next state ${\widehat{x}_{k+1}^{+}}$ are calculated. Specifically, 
 \begin{equation}
\widehat{\mathbf{P}}_{k+1} = \mathbf{F}\widehat{\mathbf{P}}_{k}\mathbf{F}^{T}+\mathbf{Q}
\label{phat}
\end{equation}
 \begin{equation}
\mathbf{K}=\widehat{\mathbf{P}}_{k+1}\mathbf{H^{T}}(\mathbf{H}\widehat{\mathbf{P}}_{k+1}\mathbf{H^{T}}+\mathbf{R})^{-1}
\label{K}
\end{equation}
 \begin{equation}
{\widehat{x}_{k+1}^{+}} = \widehat{x}_{k} + \mathbf{K}(z_{k+1}-\, \mathbf{H}\: \widehat{x}_{k+1}) 
\label{xkhat}
\end{equation}
 \begin{equation}
\mathbf{\widehat{P}_{}}^{+}=(I-\mathbf{K}\mathbf{H})\widehat{\mathbf{P}}_{k+1} 
\label{widehatP}
\end{equation}

\noindent where $z_{k+1}$ denotes the relative measurements, $\mathbf{H}$ is the measurement matrix, $\mathbf{R}$ is the matrix consisting of the variances of the process noise vector and $\mathbf{Q}$ is the covariance matrix. Fig. \ref{fig:EKF} illustrates the EKF architecture used for the dynamic position estimation of the moving vehicle. Note that the calculation of the relative measurements using the RSS of the SOPs utilizing the multilateration and LSQ techniques has been discussed in the previous section. 
\setlength{\textfloatsep}{0pt}

\section{Proposed Framework}\label{framework}
In this section, we elaborate on the proposed relative positioning system (RPS). To do so, we first develop a framework and then use it to generate the relative coordinates. The complete algorithm of the proposed system is shown in Algorithm~\ref{alg:coordinates}.
\begin{algorithm}[h]
\caption{Relative Position Algorithm}\label{alg:coordinates}
\hspace*{\algorithmicindent} \textbf{Input:} {RSS dataset at each route point denoted as $f_{k}$}
\begin{algorithmic}[1]
\State Scan the frequency spectrum $f_{k}$ at the current position of the moving vehicle and set $T_{x}$
\Procedure{Relative Position Calculation}{}
    \State Analyze $f_{k}$ and extract the distribution moments $\theta_{k}^{l}$ 
    \State $\theta_{k}^{l}$ is used to calculate the relative distances using the path-loss model in Eq.~\eqref{eqn:PL} 
    \State Apply multilateration detailed in Eq. \eqref{dn}
    \State Perform LSQ to compute  estimated locations using  Eq. \eqref{eqn:LSQ}
\State Apply WMA on $\widehat{\mathbf{x}}^{*}$ (Eq. \eqref{WMA})
\State Estimate relative position using EKF on $x_{k_{WMA}}$
\EndProcedure
\end{algorithmic}
\end{algorithm}

For each one of the $f_{k}$ spectrum sweeps, the  mean $\theta_{k}^{l}$ along a set of frequency bands $l$ is first computed over the last set of measurements as follows:
\begin{equation}
\theta_{k}^{l}=\frac{1}{N}\sum_{n=k-N}^{k}f_n^l , \forall  l
\end{equation}

\noindent where $N=1,\ldots, N_F$ is the accumulated number of sweeps carried out. Using $\theta_{k}^{l}$, the path-loss model in Eqs. \eqref{PL0}-\eqref{eqn:PL} is employed to compute relative coordinates using first multilateration and then the LSQ method to obtain $\widehat{\mathbf{x}}^{*}$ as detailed in the algorithm above. 

Using the best relative coordinates, a weighted moving average (WMA) technique is applied to eliminate short-term fluctuations and reduce the effects of extreme values. WMA also puts more weight on the most recent observations compared to the initial ones, leading to better localization accuracy. Thus, using the WMA technique the relative coordinates become more robust and the distances between reference points are better approximated. The moving average equation utilized is given as follows:
\begin{equation}
x_{k_{WMA}}=\frac{W_{i,k+1}x_{k+1}+....+W_{i,k}x_{k}}{S_{w}}
\label{WMA}
\end{equation}
where $S_{w}$ expresses the accumulated temporal weight and $x_{k_{WMA}}$ denotes the current state of the moving vehicle after the implementation of the WMA method with weights $W_{i,k}$. 

The final step of the proposed framework is related to the extraction of the optimized relative coordinates utilizing an EKF. In order to achieve an optimal trajectory, a measurement and a motion model for our moving vehicle are utilized, with the $\widehat{\mathbf{x}}^{*}$ measurements being used as landmarks in the measurement model:
\begin{equation}
x_{k+1}= x_{k}+\begin{bmatrix}
1\\ T_{s}
\end{bmatrix}u_{k}+w_{k}
\label{xk+1}
\end{equation}

\begin{equation}
d_{k+1}=\sqrt{(xl-x)^{2}+(yl-y)^{2}}+n_{k}
\label{dk+1}
\end{equation}
where $x_{k}=[x \; \;y]^{T}$ denotes the current vehicle position in 2D and $x_{k+1}=f(x_{k},u_{k})$ denotes the next estimated position calculated using the motion model with $ \mathbf{F}=\frac{\partial f}{\partial x_{k+1}}\mid _{x_{k},\; u_{k}}$ and the measurement model $d_{k+1}=h(x_{k},n_{k})$ with $\mathbf{H}=\frac{\partial h}{\partial x_{k+1}}\mid _{x_{k}}$. $u_{k}$ denotes the velocity readings used as an input to the motion model, while $w_{k}$ is the process noise with zero mean normal distribution and covariance $\mathbf{Q} = (0.1)\mathbf{I}_{2x2}$. The measurement model relates to the range between the route points with $xl$ representing the landmarks extracted using the RPS and $n_{k}$ the zero mean measurement noise with constant covariance $\mathbf{R}=0.01$. In the prediction step, the velocity readings and the motion model are utilized in order to produce a state at a given timestep $k$. Using the measurement model and importing the range measurements $z_{k+1}$ as shown in Eqs.~\eqref{K}-\eqref{widehatP}, the correction step takes place and an optimal relative trajectory is created. It should be noted that the range measurements $z_{k+1}$ are extracted using the landmarks from the previous steps of RPS.  The relative coordinates using the EKF are then calculated as follows:
\begin{equation}
x_{k+1}=x_{k}+\mathbf{K}(z_{k+1}-d_{k+1})
\label{xk+12}
\end{equation}
\begin{equation}
\widehat{X}^{*}=[\; x_{k+1}^{1},\; x_{k+1}^{2},\; ...,\;x_{k+1}^{F}]
\label{widehatX}
\end{equation}

\section{Experimental Results}\label{results}
\begin{figure}[h!]
\begin{center}
 \setlength\abovecaptionskip{-0.6\baselineskip}
 \setlength\belowcaptionskip{-0.4 pt}
\includegraphics[width=9cm]{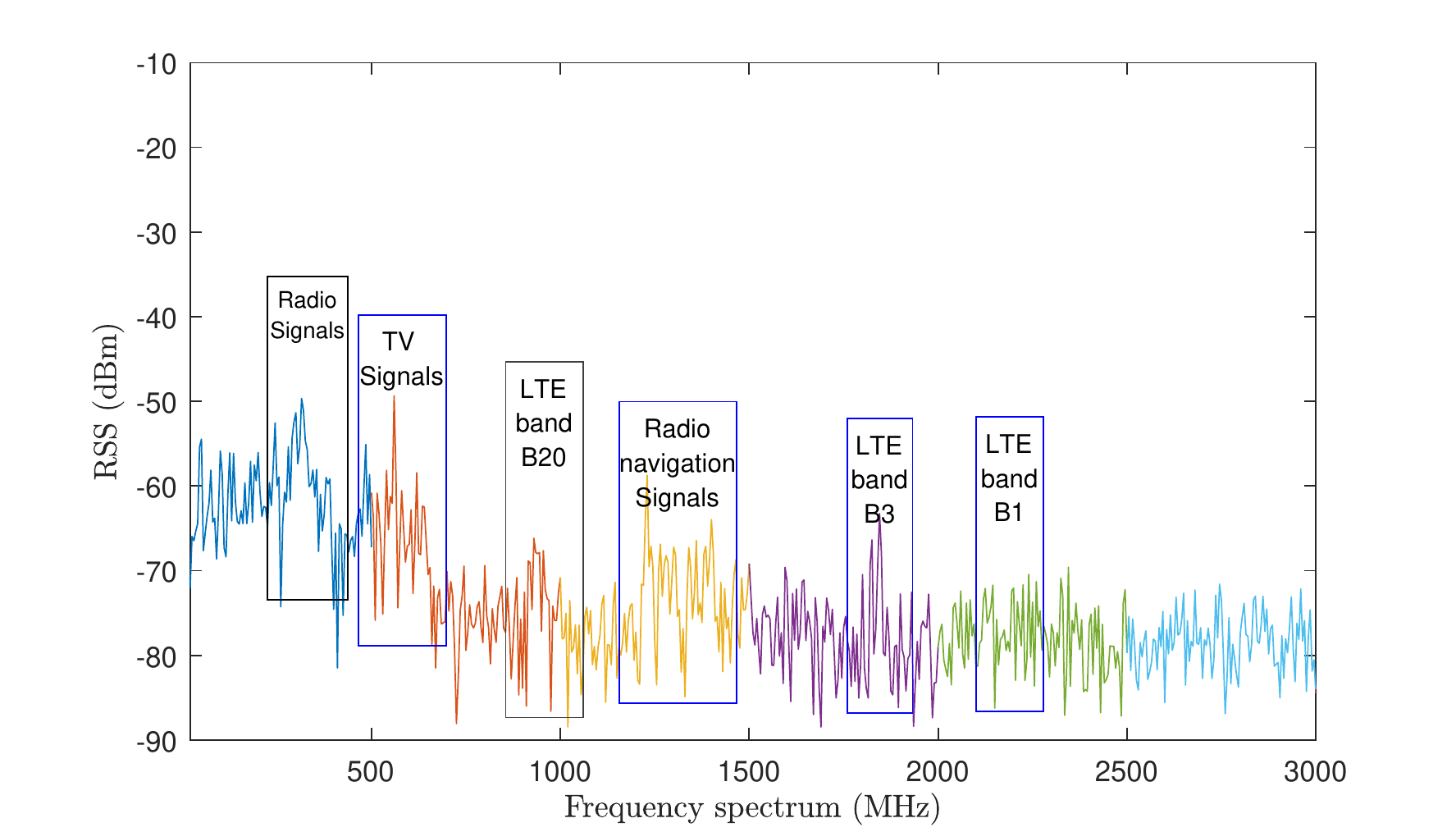}
\caption{RSS sample in the frequency range of 0-3000 MHz.}
    \label{fig:0-300}
\end{center}
\end{figure}

\begin{figure}[h]
\begin{center}
 \setlength\abovecaptionskip{-0.6\baselineskip}
 \setlength\belowcaptionskip{-0.4 pt}
\includegraphics[width=9cm]{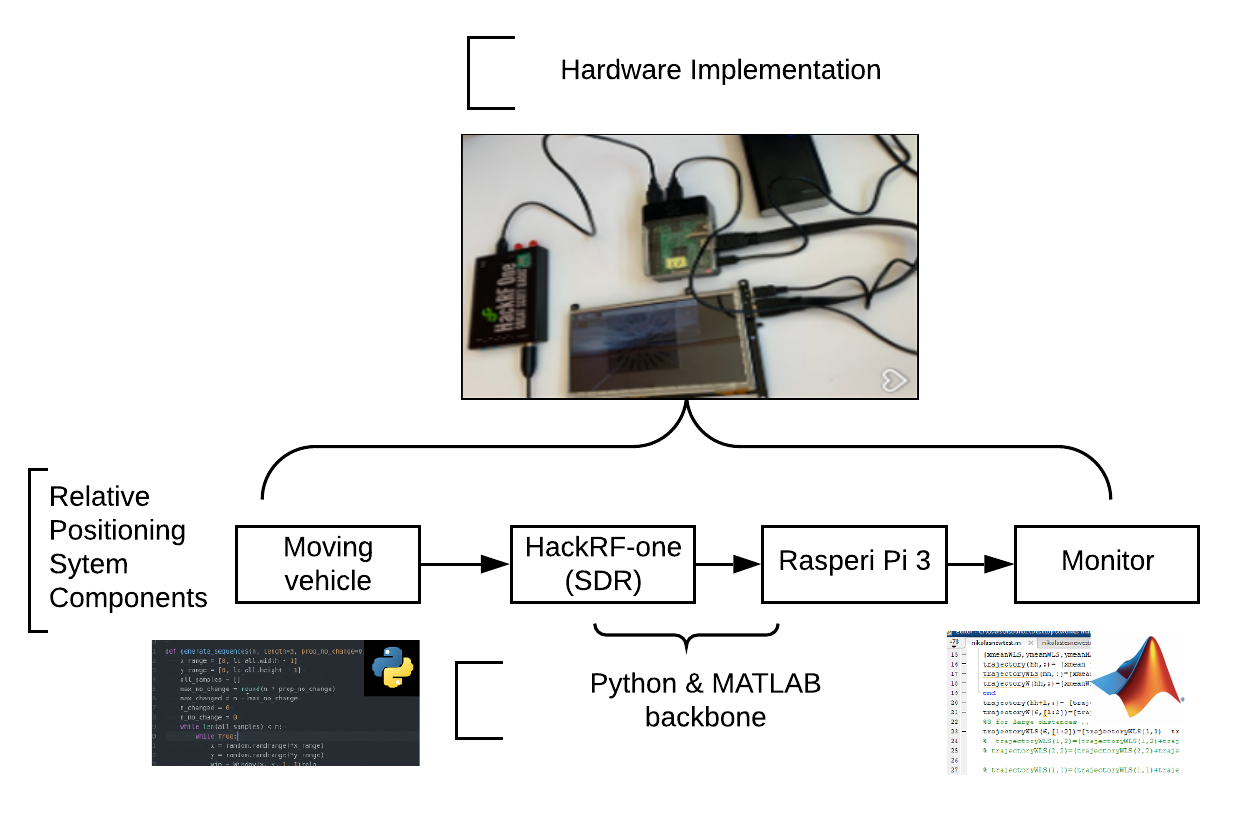}
\caption{Hardware and software set-up of the RPS.}
    \label{fig:hardware}
\end{center}
\end{figure}

A field experiment was carried out using various transmission bands and utilizing the proposed RPS in order to validate our approach. One broadband antenna was mounted on a SDR module inside a moving car to acquire SOPs over a large frequency spectrum (via HackRF-one software-defined radio) and track multiple cellular transmitters. A sample of the RSS information collected is illustrated in Fig.~\ref{fig:0-300}, while  Fig.~\ref{fig:hardware} illustrates the hardware and software set-up.

\setlength{\floatsep}{0pt}

\begin{figure}[h]
\begin{center}
 \setlength\abovecaptionskip{-0.6\baselineskip}
 \setlength\belowcaptionskip{-0pt}
\includegraphics[width=8cm]{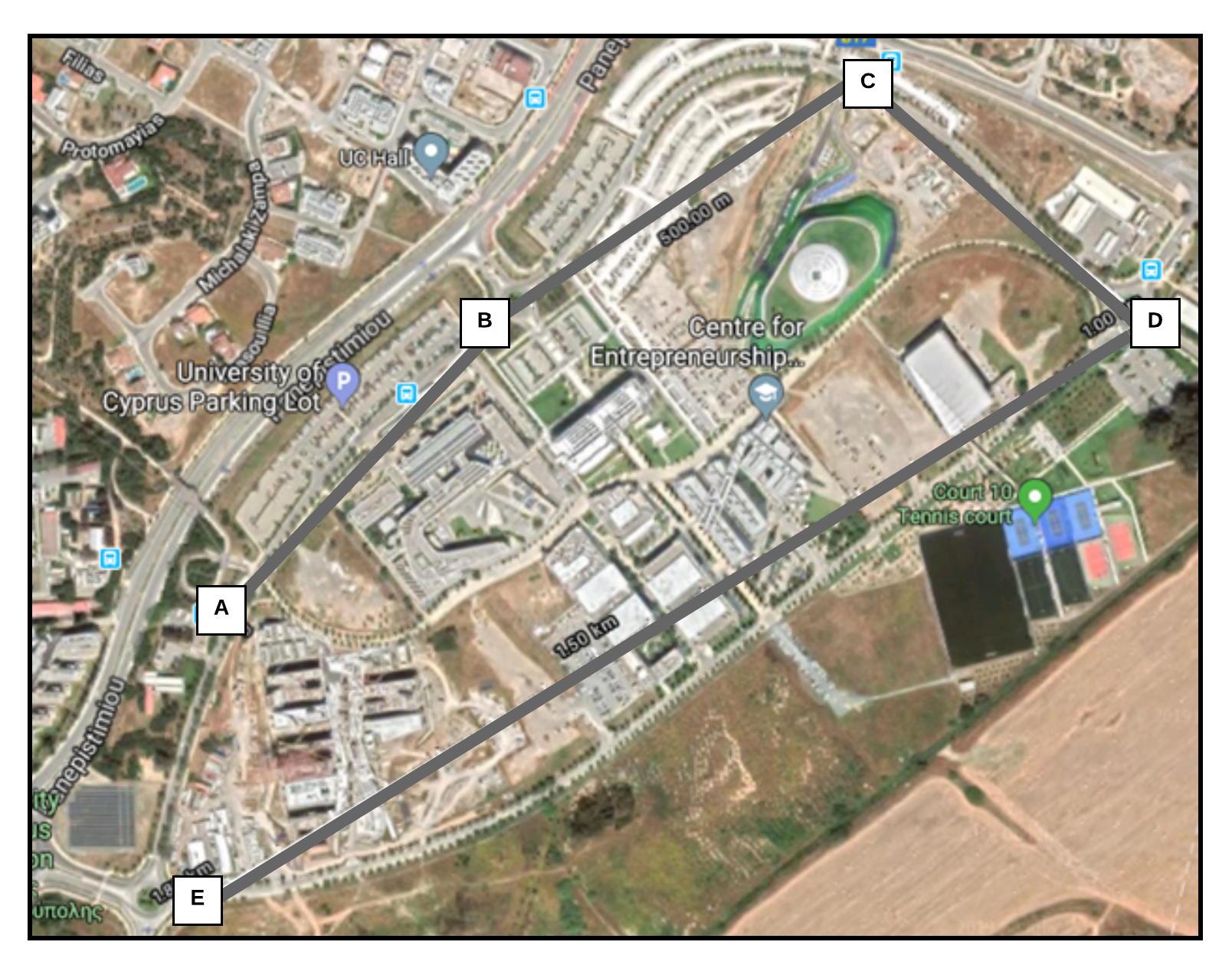}
\caption{Experimental environment and navigation results using GNSS.}
    \label{fig:GNSS}
\end{center}
\end{figure}

\begin{figure}[h]
\begin{center}
 \setlength\abovecaptionskip{-0.4\baselineskip}
 \setlength\belowcaptionskip{-0 pt}
\includegraphics[width=9cm]{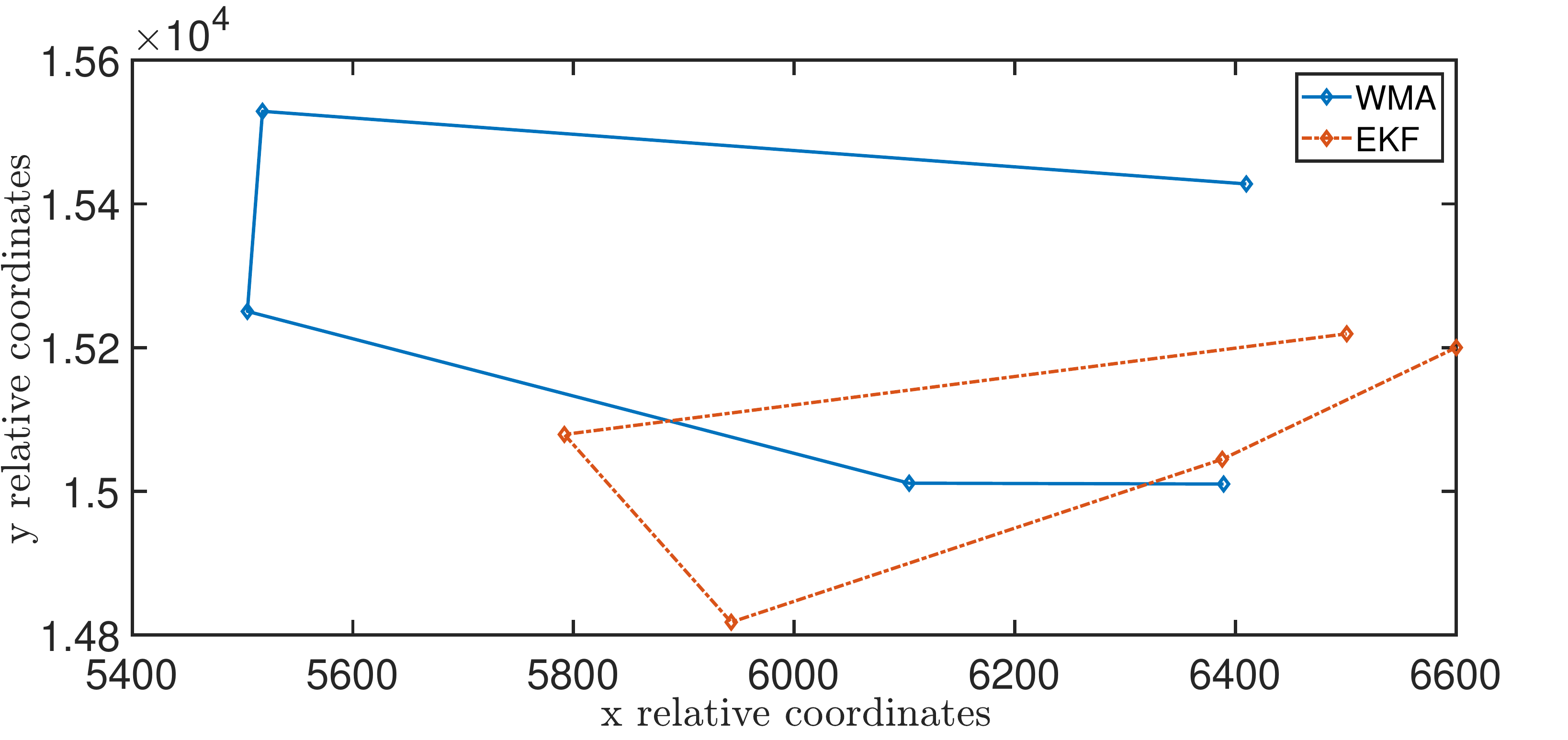}
\caption{Relative navigation results using SOPs and the WMA method, as well as the combined WMA and EKF methods.}
    \label{fig:navigation}
\end{center}
\end{figure}

Figure~\ref{fig:GNSS} illustrates the route followed by the moving vehicle and mapped using the GNSS signals, while  Fig.~\ref{fig:navigation} presents the relative trajectory extracted using the proposed system (RPS). From Fig.~\ref{fig:navigation} it is easy to deduce that by utilizing the information of the SOPs, the proposed system can track the vehicle's route and create a relative trajectory with distances between the five points analogous to the real distances shown in Fig.~\ref{fig:GNSS}. However, as expected, there is still an error in terms of distances in comparison to the real trajectory extracted using GNSS.  Thus, as previously mentioned, an EKF is also utilized to achieve better localization performance. In this case, the vehicle dynamic model and the RSS values gathered, using HackRF-ONE, are combined, using an EKF, in order to minimize the error in terms of distances between the route points. It is indeed  observed that using the EKF the error (in meters) compared to the real data extracted using the GNSS is reduced. 

\setlength{\floatsep}{-6pt}

\begin{table*}[t]
\begin{center}
\caption{Experimental Trajectory Distances.}
\begin{tabular}{|l|l|l|l|l|}
\hline
\textbf{Trajectory Points} & \textbf{A-B(m)/Difference (\%)} & \textbf{B-C(m)/Difference (\%)} & \textbf{C-D(m)/Difference (\%)} & \textbf{D-E(m)/Difference (\%)} \\ \hline
GNSS distance              & 270/0                    & 490/0                   & 260/0                   & 840/0                   \\ \hline
Relative distance - (WMA)  & 248/8.14                 & 567/15.7                & 279/7.3                 & 787/6.3                 \\ \hline
Relative distance - (EKF)    & 263/3.7                  & 500/2.04                & 290/11.5                & 800/4.75                \\ \hline
\multicolumn{5}{|l|}{RPS configuration  ($T_{x}=6$ , $n_{PL}=2.8$ and window size of WMA $=3$)}                                                                         \\ \hline

\end{tabular}
\end{center}
\label{tab:trajectorydistances}
\end{table*}

The differences between the real and the relative trajectory are tabulated in Table I.  Following the relative trajectory extraction, a number of experiments were performed to ascertain the performance when different parameters were altered (e.g., the number of transmitters and the path-loss exponent coefficient). Initially, a number of $6$ relative transmitters ($T_{x}$) were assumed to be randomly located in space using a frequency range of $0-3500$ MHz and the results of the proposed RPS technique are illustrated in Fig.~\ref{fig:navigation}. In addition, a second experiment was performed using $13$ transmitters. It was observed that utilizing $13$ $T_{x}$ the trajectory could again be re-created, but the error in distance between the five points became larger. This demonstrates the influence that different frequency bands can have on positioning and the need for proper selection of those to be employed.

In the second part of the experimental analysis, higher frequency bands were utilized for a system of $6$, $9$, and $13$ $T_{x}$, respectively. It was observed that higher frequencies cannot be used for tracking purposes, as the relative trajectory deteriorated leading to orientation displacement. Clearly, higher frequency bands deteriorate faster over distance and thus are not favorable for outdoor fast-moving scenarios. 

Additionally, experiments with different path-loss exponent coefficients ($n_{PL}$) were performed. Analyzing the results obtained, it is clear that the orientation of the relative trajectory is not affected by the values of $n_{PL}$. However, it is observed that the distances between the five points differ and are depended on the different values of $n_{PL}$. The different RPS configurations examined and the results obtained are tabulated in Table II. 

\begin{table*}[t]
\begin{center}
\caption{Experimental analysis of RPS parameters}
\begin{tabular}{|c|c|l|c|c|c|c|}
\hline
\multicolumn{3}{|l|}{\textbf{Trajectory Points}}                                                                            & \multicolumn{1}{l|}{\textbf{A-B(m) -  $T_{x}$ (6/9/13) }} & \multicolumn{1}{l|}{\textbf{B-C(m) - $T_{x}$  (6/9/13) }} & \multicolumn{1}{l|}{\textbf{C-D(m) - $T_{x}$  (6/9/13) }} & \multicolumn{1}{l|}{\textbf{D-E(m) - $T_{x}$  (6/9/13) }} \\ \hline
\multirow{3}{*}{\textbf{$n_{PL}$}} & \textit{\textbf{2.8}}  & \multicolumn{1}{c|}{\multirow{3}{*}{\textbf{WMA window size (3)}}} & 248/\;367\;/255                                            & 567/\;600\;/470                                            & 279/\;454\;/313                                            & 787/\;950\;/740                                            \\ \cline{2-2} \cline{4-7} 
                              & \textit{\textbf{2.85}} & \multicolumn{1}{c|}{}                                              & 291/\;259\;/176                                            & 655/\;515\;/327                                            & 289/\;400\;/217                                            & 916/\;676\;/650                                            \\ \cline{2-2} \cline{4-7} 
                              & \textit{\textbf{2.9}}  & \multicolumn{1}{c|}{}                                              & 200/\;185\;/140                                            & 460/\;436\;/300                                            & 210/\;287\;/170                                            & 650/\;500\;/550                                            \\ \hline
\multirow{3}{*}{\textbf{$n_{PL}$}} & \textit{\textbf{2.8}}  & \multirow{3}{*}{\textbf{WMA window size (4)}}                      & 336/\;342\;/255                                            & 325/\;344\;/445                                            & 419/\;525\;/255                                            & 719/\;425\;/500                                            \\ \cline{2-2} \cline{4-7} 
                              & \textit{\textbf{2.85}} &                                                                    & 291/\;260\;/200                                            & 229/\;270\;/220                                            & 393/\;456\;/309                                            & 580/\;342\;/424                                            \\ \cline{2-2} \cline{4-7} 
                              & \textit{\textbf{2.9}}  &                                                                    & 200/\;190\;/170                                            & 224/\;188\;/210                                            & 360/\;400\;/257                                            & 524/\;315\;/400                                            \\ \hline
\multicolumn{7}{|l|}{\begin{tabular}[c]{@{}l@{}}$n_{PL}$: path-loss exponent, $T_{x}$: number of relative transmitters\end{tabular}}         \\ \hline

\end{tabular}
\end{center}
\label{tab:RCSparameters}
\end{table*}

\begin{figure}[h!]
\begin{center}
 \setlength\abovecaptionskip{-0.6\baselineskip}
 \setlength\belowcaptionskip{-0pt}
\includegraphics[width=9cm]{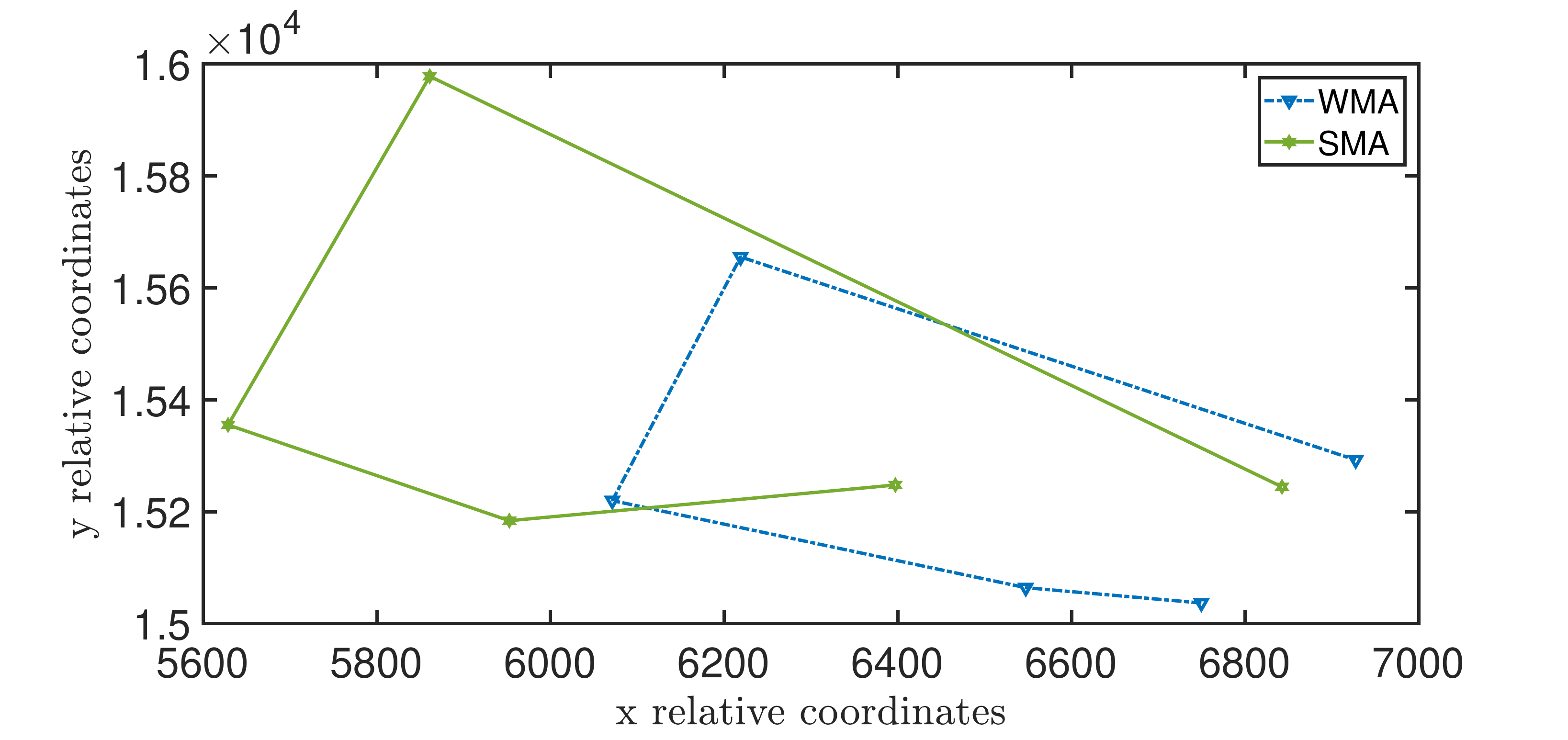}
\caption{Relative coordinates using different moving average techniques.}
    \label{fig:WMAANDSMA}
\end{center}
\end{figure}

Further, a comparison between various moving average filters was performed (Fig.~\ref{fig:WMAANDSMA}) demonstrating that the WMA approach obtains better results compared to the single moving average (SMA) technique. This of course relates to the higher value placed on the calculations made with a larger set of values which improves on the estimated position obtained from RSS data.

\setlength{\textfloatsep}{0pt}

\begin{figure}[h]
\begin{center}
 \setlength\abovecaptionskip{0.5\baselineskip}
 \setlength\belowcaptionskip{-0.4 pt}
\includegraphics[width=9cm,,height=6cm]{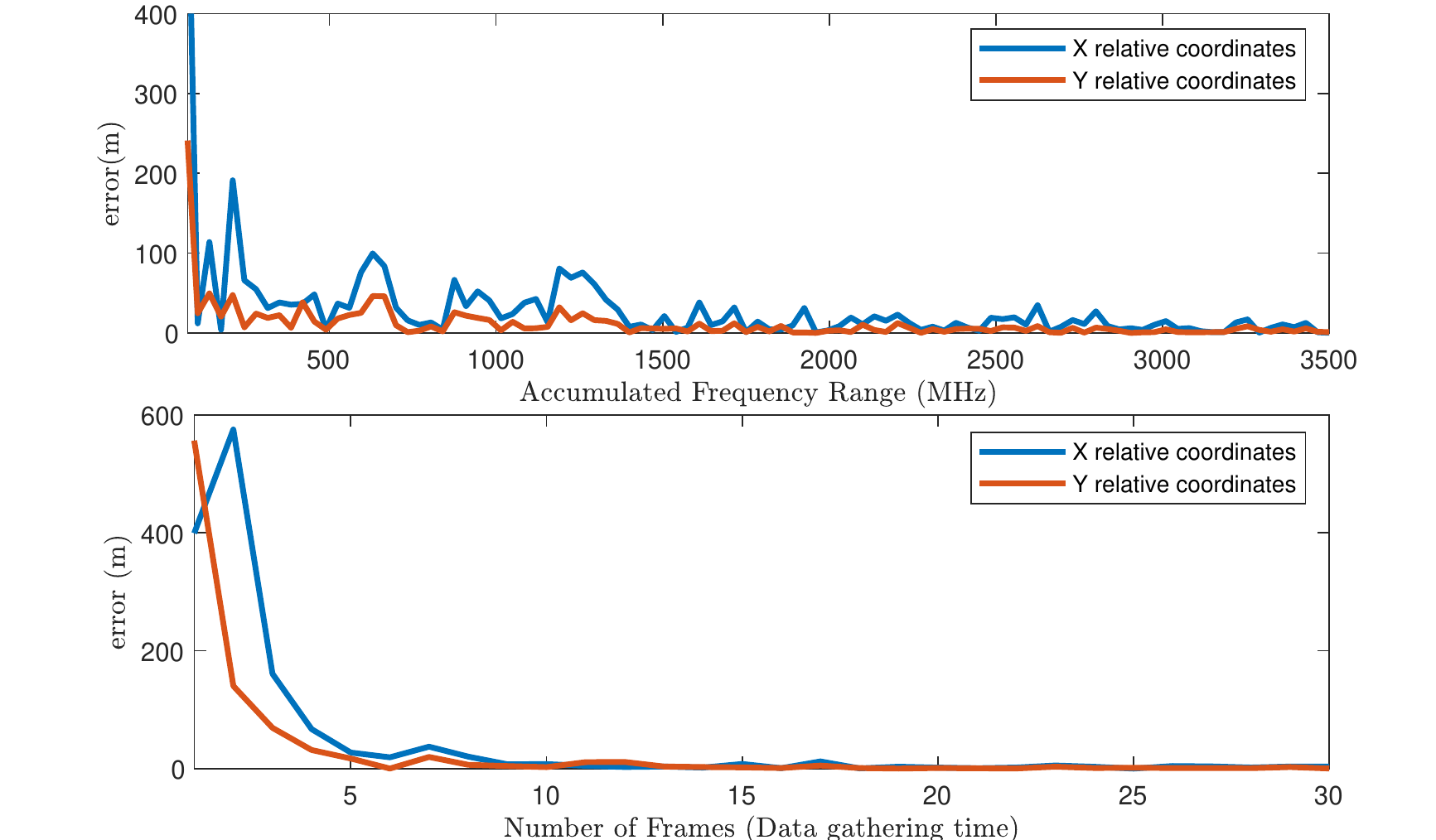}
\caption{Relative coordinates convergence over frequency  range (top); relative coordinates convergence over time (bottom).}
\label{fig:convergence}
\end{center}
\end{figure}

Finally, it should be noted that in order to achieve the aforementioned results, information in the frequency range of  $0-3500$ MHz was utilized, leading to fast relative coordinate convergence (Fig.~\ref{fig:convergence} (top)). Clearly, as we include more data, utilizing the information of a larger spectrum, the error (variation of relative coordinates - in meters) decreases. Another factor affecting the experimental results is the data collection time. As shown in Fig.~\ref{fig:convergence} (bottom), as the time increases the error is reduced (the relative coordinates converge to a point), providing better localization results. The trade-off in this case, as expected, is the increase in the data processing time.

\setlength{\textfloatsep}{0pt}

\section{Conclusions}\label{conclusions}
In this work, an innovative and low-cost navigation approach (RPS) is proposed, exploiting the information from SOPs in the environment of a moving vehicle. The proposed technique does not require a  priori knowledge of the location of the transmitters or the initial position of the moving vehicle in order to extract a relative trajectory. A number of experiments were performed to validate the proposed approach, varying the number of transmitters, the path-loss exponent, and the frequency range to investigate how they affect the performance of the proposed technique. It was further demonstrated that the utilization of EKF leads to better accuracy in terms of vehicle localization. 
Ongoing work includes using additional sensor data (e.g., IMU information) and event-based techniques for more efficient navigation using SOPs. 

\section*{Acknowledgment}
This work has been supported by the European Union's Horizon 2020 research and innovation programme under grant agreement No 833611 (CARAMEL) and 739551 (KIOS CoE) and from the Government of the Republic of Cyprus through the Directorate General for European Programmes, Coordination and Development.

\end{document}